\documentclass{elsart}
\usepackage{epsfig}
\usepackage{rotate}
\usepackage{amssymb}
\usepackage{amsmath}

\begin{document}


\begin{frontmatter}
\begin{center}
{\large \bf The muon anomalous magnetic moment and a new light gauge boson.} 
\end{center}
\vspace{0.5cm}

\begin{center}  
S.N.~Gninenko\footnote{E-mail address:
 Sergei.Gninenko\char 64 cern.ch} and 
N.V.~Krasnikov\footnote{Nikolai.Krasnikov\char 64 cern.ch}\\
{\it Institute for Nuclear Research of the Russian Academy of Sciences,\\ 
Moscow 117312}
\end{center}                                                            

\begin{abstract} 
It is shown that the 2.6 $\sigma$ discrepancy between the predicted and 
recently  measured value of the 
anomalous  magnetic moment of positive muons  could be explained by 
the existence of a new light  gauge boson X with a mass $M_X \leq O(5)~GeV$.\
Phenomenological bounds on the X coupling  are discussed.
\end{abstract}
\end{frontmatter}

\vspace{1.0cm}

The recent precise measurement of the  anomalous magnetic 
moment of the positive muon $a _{\mu}= (g-2)/2$ from Brookhaven AGS experiment 821 \cite{exp} gives result which is about
$2.6 \sigma$ higher than the 
Standard Model prediction
\begin{equation}
a^{exp}_{\mu} -a^{SM}_{\mu} = (43 \pm 16)\times 10^{-10}
\end{equation}
This result may signals the existence of new physics beyond the 
Standard Model. 
At present the standard explanation of this result suggested in a few recently
appeared papers is the supersymmetry with 
the chargino and sneutrino lighter than 800 GeV \cite{theor2}. 
Note also possible explanations related with existence of leptoquarks 
\cite{theor3} 
or existence of some exotic flavour-changing interactions \cite{theor4}. 
All these explanations assume the existence of new particles with masses 
$\geq O(100)~GeV$.

In this note  we suggest that the recent BNL result gives an evidence for the 
existence of the physics with new light particle. As an example of the 
realization of this scenario we consider the model with       
light gauge boson X ($M_X \leq O(5)~GeV$) and briefly 
discuss possible phenomenological implications. 

To be concrete besides standard $SU_c(3) \otimes SU(2)_L \otimes 
U(1)_Y$ gauge group  
consider additional $U(1)_X$ interaction which commutes with 
standard gauge group. In other words our model has 
$SU_c(3) \otimes SU(2)_L \otimes U(1)_Y \otimes U(1)_X $ 
gauge group.
We assume that all quarks have the same interaction with new gauge group 
$U(1)_X$ but the interactions of each lepton flavour with new X-boson could 
be different. In such model the standard Higgs doublet has zero X-charge 
so there is no mixing between X-boson and Z-boson. The interaction of the X-boson 
with quarks and leptons can be written in the form 
\begin{equation}
 L_X = g_X[Q_{BX}B^{\alpha} + Q_{eX}L^{\alpha}_e +   
Q_{\mu X}L^{\alpha}_{\mu} + 
Q_{\tau X}L^{\alpha}_{\tau}]X_{\alpha}
\end{equation}
where $ B^{\alpha} = \sum_{q =u,d,s,...}\bar{q}\gamma^{\alpha}q$, 
$L^{\alpha}_e = \bar{e}\gamma^{\alpha}e + \bar{\nu}_{eL}\gamma^{\alpha}
\nu_{eL}$,... . From the requirement of the absence of 
the $\gamma_5$ anomalies 
an additional constraint $3Q_{BX} + Q_{eX} + Q_{\mu X} + Q_{\tau X} = 0$ 
arises. The X-boson gives additional contribution to the anomalous 
magnetic moment of lepton
\begin{equation}
\delta a_l = \frac{Q^2_{lX}\alpha_X}{\pi}\int^1_0 \frac{x^2(1-x)}
{x^2 + (1-x)M^2_X/m^2_l},
\end{equation}       
where $\alpha_X = g^2_X/4\pi$ and $M_X$ is the mass of the X-boson.
We shall use the normalisation  $Q_{\mu X} = 1$. 
For 
$M_X \ll m_{\mu}$ we find from Eq.(1) that
\begin{equation} 
\alpha_{X} =  (2.7 \pm 1) \times 10^{-8}
\end{equation}
For another limiting case $M_X \gg m_{\mu}$  Eq.(1) leads to  
\begin{equation}
\alpha_X\frac {m^2_{\mu}}{M^2_X} = (4.1 \pm 1.5)\times 10^{-8}
\end{equation}
To suppress the contribution of the X-boson 
to the anomalous magnetic moment of the electron we assume 
that \footnote{Experimental data on $\nu_{\mu}(\bar{\nu}_{\mu})e$ scattering 
lead to $|Q_{eX}| \ll 1$, see Eq.(8).} 
$M_{X} \geq 10~MeV$ or $Q_{eX} << Q_{\mu X}$.  

The phenomenology of the light X-boson has been studied in 
refs.\cite{eric}-\cite{gnin} (see also \cite{okun}) 
where untrivial bounds on the interaction of the X-boson with quarks and 
leptons have been derived.\\

Consider first possible manifestations of the X-boson in 
{\it neutrino reactions}. 
An account of the X-boson exchange between muon neutrino, electrons and 
quarks leads to the additional four-fermion interaction
\begin{equation}
\delta L_{4} = \frac{g^2_X}{m^2_X -t} \bar{\nu}_{\mu L}
\gamma^{\mu}\nu_{\mu L}
(Q_{eX}\bar{e}\gamma_{\mu} e + Q_{BX}\sum_{i} \bar{q}_i\gamma_{\mu}q_i)
\end{equation}
Here, t is the square of the momentum transfer.
Consider $\nu_{\mu}(\bar{\nu}_{\mu})e$-scattering.
The typical cutoff on the momentum transfer in neutrino-electron scattering is 
$t_0\simeq -400~MeV^2$ and the cross section is maximal for minimal momentum 
transfer. The experimental value of the effective vector coupling of the 
electron with muon neutrino $g^{\nu e}_{V} = 0.3009 \pm .0015$ \cite{pdg}.
Comparison of the contribution from the effective interaction $\delta L_4$ 
to the $\nu_{\mu}-e$ scattering with the experimental data \cite{pdg}
leads to the bound \cite{dobr}
\begin{equation}
|Q_{eX}|g^2_{X} \leq (M^2_X +400~MeV^2)\times 5\cdot10^{-7}~GeV^{-2}
\end{equation}
For masses $M_{X} \ll 20~MeV$ we find that 
\begin{equation}
|Q_{eX}| \leq 2.3 \cdot 10^{-3}
\end{equation}
For $M_X \gg 20~MeV$ we obtain that
\begin{equation}
|Q_{eX}| \leq 0.04
\end{equation}
 Analysis of 
neutral current $\nu$ reactions on nuclei allow to restrict value of $|Q_B|$.
For deep inelastic $\nu_{\mu}N$- scattering (DIS) with the minimal momentum 
transfer $t_0 = -10~GeV^2$ (for this value we are sure that perturbative 
QCD works rather well) the limit on the discrepancy between the 
 predicted and measured values of DIS cross section could be evaluated to 
be  $\lesssim 3~\%$ \cite{pdg}.\ 
This results in the bound for X-boson contribution 
\begin{equation}
g^2_X|Q_{BX}| \leq 5\cdot10^{-6}
\end{equation}
  and gives
\begin{equation}
|Q_{BX}| \leq 60~(M_X \ll m_{\mu})
\end{equation}
\begin{equation}
|Q_{BX}| \leq 40 m^2_{\mu}/M^2_X~(M_X \gg m_{\mu})
\end{equation}
For coherent pion production in neutrino and antineutrino scattering on nuclei 
\cite{kopel} the typical 
minimal momentum transfer is $t_0\simeq -0.02~GeV^2$. From the assumption
that X-boson contribution is $\lesssim 25~\%$, which is taken as a limit 
on the difference between the experimental 
and theoretical values of the cross section \cite{kopel}, one can find that 
\begin{equation}
|Q_{BX}| \leq O(1)~(M_X \ll m_{\mu})
\end{equation}
\begin{equation}
|Q_{BX}| \leq O(0.4)~(M_X \gg m_{\mu})
\end{equation}
Similar to neutrino scattering experiments possible existence of the X-boson
could manifests itself in {\it rare meson decays}.  
If the X-boson is lighter than $\pi^0$-meson it is 
possible to search for it in the decay $ \pi^{0} \rightarrow \gamma + X $ 
 \cite{dobr,gnin}.  The branching ratio of 
the $\pi^{0} \rightarrow \gamma + X$ is \cite{dobr}
\begin{equation}
Br(\pi^{0} \rightarrow X + \gamma) = \frac{18Q^2_{BX}\alpha_X}{\alpha}
(1- \frac{M^2_X}{m^2_{\pi}})^3
\end{equation}
From the experimental bound on the branching ratio 
$Br(\pi^{0} \rightarrow X + \gamma)\lesssim 3\cdot10^{-5}$ \cite{nomad} 
one can find that 
\begin{equation}
|Q_{BX}| \leq 2.1 
\end{equation}
for $m_X \ll m_{\mu}$. The decay $K \rightarrow 
\pi + X$ has been studied in ref. \cite{aliev}. The decay width 
$\Gamma(K \rightarrow \pi + X)$ is proportional to 
$M^2_X/m^2_{\pi}$ and it is maximal for 
$M_X \approx 200~MeV$ \cite{aliev}. It appears that  bound on $|Q_{BX}|$
from the existing experimental data  is  weaker than the 
corresponding bounds coming from the neutrino-nuclei scattering.

The  X-boson contribution to the rare decays $K\rightarrow
\pi\mu\bar{\mu}, \pi \nu\bar{\nu}$ comes from the electromagnetic 
penguin diagram \cite{buras} with the replacement 
of the photon exchange on the X-boson exchange.
The corresponding flavour changing $\bar{s}dX$ effective interaction 
in full analogy with the case of the electromagnetic penguin \cite{buras} is 
determined by the effective vertex  
\begin{equation}
\bar{s}\gamma d = -i\lambda_{i} \frac{G_F}{\sqrt{2}}\cdot 
\frac{g_XQ_{BX}}{8\pi^{2}}\cdot D_0^{'}(x_i)\bar{s}(q^2\gamma_{\mu} - q_{\mu}
\hat{q})(1-\gamma_{5})d \,,
\end{equation}
where $\lambda_i = V^{*}_{is}V_{id}$, $x_i = m^2_{q_i}/m^2_w$ 
and $D^{'}_0(x_i) = O(1)$. Here, $q_{\mu}$ is the outgoing X-boson momentum. 
Using the effective vertex of Eq.(17) one can find that the contribution 
of the X-boson exchange to the $K^{+} \rightarrow \pi^{+}\nu \bar{\nu}$ 
effective Hamiltonian is 
\begin{equation}
\delta H_{eff}(K^{+} \rightarrow \pi^{+}\nu\bar{\nu}) \sim 
\frac{G_F}{\sqrt{2}}\lambda_t \frac{\alpha_X Q_{BX}}{2\pi}
\frac{(p_{\nu} +p_{\bar{\nu}})^2}{(p_{\nu} + p_{\bar{\nu}})^2 - m^2_X}
(\bar{s}d)_{V-A}(\bar{\nu}\nu)_{V-A}
\end{equation}
and it is much smaller (by a factor $O(10^{-6})Q_{BX})$ than the standard 
model contribution 
\begin{equation}   
H_{eff}(K^{+} \rightarrow \pi^{+}\nu\bar{\nu}) \sim 
\frac{G_F}{\sqrt{2}}\lambda_{t} \frac{\alpha}{2\pi \sin^2 \theta_w}
(\bar{s}d)_{V-A}(\bar{\nu}\nu)_{V-A}
\end{equation}
to this decay. The similar situation takes place for the other 
$K \rightarrow \pi \mu \bar{\mu}$, $B \rightarrow X_{s,d} \nu \bar{\nu}$ 
decay modes.\ Finally, we conclude that there is no 
stringent bounds on $|Q_{BX}|$ from the rare meson decays.  

Finally consider bounds on  X-boson couplings with electron and
 muon from the spectroscopy measurement of the $1s-2s$ interval 
in  {\it muonium $(\mu^{+} e^{-})$ atoms} \cite{muonium}. 
The result of this measurement can be 
interpreted as a measurement of the muon-electron charge ratio \cite{muonium}
\begin{equation}
\frac{Q_{\mu^{+}}}{Q_{e^{-}}} = -1 -(1.1 \pm 2.1)\cdot10^{-9}
\end{equation}
The additional interaction of muon with electron due to the X-boson exchange 
leads to the additional (in comparison with standard Coulomb potential) 
non-relativistic interaction between electron and $\mu^{+}$ meson
\begin{equation}
\delta V(r) = -\frac{\alpha_{X}Q_{eX}}{r}exp(-M_Xr)
\end{equation}
For very light X-boson ($M_X \ll m_e\alpha$) the exponential factor in Eq.(21) 
is negligible and  an account of the additional interaction of Eq.(21) 
is equivalent 
to the standard Coulomb  interaction of the electron with $\mu^{+}$ with 
muon electric charge equal to $Q_{\mu^{+}} = 
1 + \alpha_{X}/\alpha \cdot Q_{eX}$. So for $M_X \ll m_e\alpha$ we find 
from Eq.(4) and Eq.(20) that 
\begin{equation}
Q_{eX} = (3  \pm 5.7) \cdot10^{-4}
\end{equation}
For $M_{X} \gg m_e\alpha$ we have exponential damping 
$exp(-M_{X}/m_e\alpha)$ for 
the X-boson exchange contribution to the $1s-2s$ energy difference of 
muonium and there is no useful bound on $Q_{eX}$ from the muonium 
spectroscopy.\ 
Note that the spectroscopy of the hydrogen atoms is not so ``clean'' 
from the theoretical point of view as the muonium one due to composite 
nature of proton  that prevents  the extraction of 
stringent  bounds on 
the interaction of the X-boson with quarks.

To summarise,  the existing experimental data give the most stringent 
bound on $|Q_{eX}| \ll 1$ whereas the bound on $|Q_{XB}|$ is rather weak. 
  
Especially interesting (in particular, from the aesthetic point of view) 
 is the case when X-boson interacts only with second 
and third lepton generations ( $ Q_{BX} = Q_{eX} = 0, Q_{\mu X} = -
Q_{\tau X} = 1$) that makes the search for such particle quite difficult.\
 For $M_X \leq 2m_{\mu}$ the X-boson decays mainly to 
$X \rightarrow \nu_{\mu}
\bar{\nu}_{\mu}, \nu_{\tau}\bar{\nu}_{\tau}$. For $2m_{\mu} <M_X <2m_{\tau}$ 
the additional decay mode to $\mu^{+}\mu^{-}$ opens up with the branching
ratio $\sim 0.5$ 
if $2m_{\mu} \ll M_{X}$.\ For $M_{X} > 2 m_{\tau}$ the  
decay channel $X \rightarrow \tau \bar{\tau}$ is also opened.\  

The best limit comes  
from the non-observation of such X-boson at LEP1 in the 
decay $ Z \rightarrow \mu^{+} \mu^{-} + X, \tau^{+} \tau^{-} + X$. We 
obtain  $\alpha_{X} \leq O(10^{-4})$ that leads to the 
bound 
\begin{equation}
M_X \leq O(5)~GeV
\end{equation}
 on the X-boson mass. Note that bound (23) holds only if the X-boson explains 
the $g-2$ anomaly.\ 
It is interesting that the existence of such gauge 
interaction prohibits diagonal neutrino Majorana masses $\nu_{\mu}\nu_{\mu}$ 
and $\nu_{\tau}\nu_{\tau}$ and allows non diagonal mass 
term $\nu_{\tau}\nu_{\mu}$ before $U(1)_X$ gauge symmetry breaking. After 
$U(1)_X$ gauge symmetry breaking small diagonal terms appear that leads 
to the maximal 
$\nu_{\mu} -\nu_{\tau}$ mixing \footnote{We are indebted to Prof. S.L. Glashow 
for interesting comments on this subject.} with the typical prediction
$\Delta m^2_{\nu_{\mu}\nu_{\tau}} \ll m^2_{\nu_{\mu}}$. 
.
 
The branching ratio for the $\mu \rightarrow e\bar{\nu}_e\nu_{\mu}X
(X \rightarrow \nu\bar{\nu})$ is predicted to be $O(\alpha_{X}/\pi) 
\sim O(10^{-8})$ that is too small to be observable in the nearest future.\
The X-boson  could be produced in the 
reaction $\mu + N \rightarrow \mu + X + ...$ and be detected 
(for $M_{X} >2 m_{\mu}$) through its 
decay into  muon pair $X \rightarrow \mu^{+}\mu^{-}$ with the
$\sigma(\mu + N \rightarrow \mu +X +...) / \sigma(\mu + N \rightarrow 
\mu + ...) \sim O(\alpha_{X}/\pi)$.
Such reaction could be searched for, e.g. in the COMPASS experiment at CERN. 

Note that massless X-bosons could be associated with muonic(leptonic) 
photons, introduced to explain conservation of the muon(lepton) number, see 
\cite{okun} and references therein.\ 
Interestingly, the existing experimental limit on
coupling strength between the muonic photons and muons, 
$\alpha_{\mu} < (1.1 \div 2.3) \times 10^{-8}$ \cite{charm}, is in the region 
of  values required to 
explain the BNL result, $\alpha_{\mu} = (2.7\pm 1.0) \times10^{-8}$.\   

Although many models explaining the BNL discrepancy   
appeared recently, see e.g.  \cite{theor2}-\cite{theor4}, all of them 
require very high energies for their experimental tests.\footnote{For some of
the many references appeared during the completion of this work
see also \cite{stream}.}  In this note 
the discrepancy is explained by the existence of a new weakly interacting
light gauge X-boson, which is not excluded by the experimental 
data. This makes it still 
interesting for further experimental searches at present accelerators. 
 
We are grateful to Prof S.L. Glashow for his interest to our work and 
useful  remarks.\ The work of N.V.K. has been partly supported by INTAS-CERN 377 grant.


\end{document}